\documentclass[aps,prd,showkeys,showpacs,superscriptaddress,floatfix]{revtex4}

\usepackage{graphics,epsfig}
\usepackage{epstopdf}
\usepackage{subfigure}
\usepackage{palatino}
\usepackage{changes}
\usepackage[colorlinks=true,linkcolor=blue,urlcolor=blue,citecolor=blue]{hyperref}
\usepackage[toc,page]{appendix}
\usepackage[normalem]{ulem}
\usepackage{adjustbox}
\usepackage{latexsym}
\usepackage{amsmath}
\usepackage{amssymb}
\usepackage{amsfonts}
\usepackage{times}
\usepackage{graphicx}
\usepackage{dcolumn}
\usepackage{amsmath}
\usepackage{array}
\usepackage{wrapfig}
\usepackage{multirow}
\usepackage{tabularx}
\newcommand{\br}[1]{\left(#1\right)}

\newcommand{\dal}{\Box \phi}
\newcommand{\dpp}{\left(\nabla \phi \right)^2}

\renewcommand{\a}{\alpha}
\renewcommand{\b}{\beta}

\begin{document}

\title{Black hole with confining electric potential in scalar-tensor description of regularized 4-dimensional Einstein-Gauss-Bonnet gravity}

\author{A. {\"O}vg{\"u}n}
\email{ali.ovgun@emu.edu.tr}
\affiliation{Physics Department, Eastern Mediterranean University, Famagusta, 99628 North
Cyprus via Mersin 10, Turkey.}

\begin{abstract}
In this paper, we analytically present
an exact confining charged black hole solution to the scalar-tensor description of regularized 4-dimensional Einstein-Gauss-Bonnet gravity (4DEGBG) coupled to non-linear gauge theory containing $\sqrt{-F^{2}}$ term.  Part of the importance of using a non-trivial $4D$ theory of Gauss-Bonnet gravity is that it is proven the absence of the Ostrogradsky instability was formulated in [D. Glavan and C. Lin, Phys. Rev. Lett.
124, 081301 (2020) \cite{Glavan:2019inb}] and then regularized 4DEGBG theory was obtained in [Hennigar et al. JHEP 07, 027 (2020) \cite{Hennigar:2020lsl}] and [Fernandes et al. Phys. Rev. D 102, no.2, 024025 (2020) \cite{Fernandes:2020nbq}] as well as this theory gains many attention because of its important finding for black hole physics. We therefore also study
some properties of this solution such as temperature, specific heat,shadow and quasinormal modes. Our results here show that a confining charge $f$ gives a significant contribution to the shadow of the black hole as well as quasinormal modes frequencies.
\end{abstract}
\date{\today}
\keywords{Black hole; Exact solutions; Gauss-bonnet gravity; QCD-inspired confining potential; Shadow}

\pacs{95.30.Sf, 04.70.-s, 97.60.Lf, 04.50.+h}

\maketitle


\section{\label{sec:level1}Introduction}

Gravitational tests have been conducted in a variety of cosmic settings since the the 1919 solar eclipse, which is the first evidence of General Relativity (GR) \cite{Einstein:1916vd,Dyson:1920cwa}. Not long ago, gravitational tests have been organized to explore gravity outside the solar system and even on a cosmological scale. For instance, the Laser Interferometer Gravitational-Wave Observatory (LIGO) detects the gravitational waves that propagate through the fabric of spacetime as predicted by GR \cite{Abbott:2016blz}. On the other hand, the Event Horizon Telescope (EHT) collaboration, has recently imaged M87’s central black hole, and provided a number of enlightening answers \cite{Akiyama:2019cqa}. Based on an analysis of the black hole’s shadow, the team showed a solitary search of GR, boosting knowing about the properties of black holes (BHs) and eliminating many modified gravity theories \cite{Psaltis:2020lvx,Akiyama:2021qum}.

GR stands among the most famous theories ever created, but it has some long-standing weak points for instance, combining GR with quantum mechanics (quantum gravity) or singularity problem \cite{Capozziello:2011et,Demir:2021moq}. One of the important description of quantum theory for a linear confinement phenomena using the actions of effective nonlinear gauge field (NGF) was studied by ‘t Hooft \cite{tHooft:2002gdr}. It was shown that the  nonlinear terms in NGF interpret as “infrared counterterms”. Along with the Maxwell term, one can add a square root of the field strength squared, and construct a special type of non-Maxwell nonlinear effective gauge field model as given by \cite{Vasihoun:2014pha,Guendelman:2003ib,Gaete:2006xd,Korover:2009zz,Guendelman:2011sm,Guendelman:2012ve,Guendelman:2018tzi}:

\begin{equation}
\begin{array}{c} \label{ac}
S=\int d^{4} x L\left(F^{2}\right) \quad, \quad L\left(F^{2}\right)=-\frac{1}{4} F^{2}-\frac{f}{2} \sqrt{-F^{2}} \\
F^{2} \equiv F_{\mu \nu} F^{\mu \nu} \quad, \quad F_{\mu \nu}=\partial_{\mu} A_{\nu}-\partial_{\nu} A_{\mu},
\end{array}
\end{equation}
where $f$ being a positive coupling constant. Moreover, the square root of the Maxwell term is appeared naturally as a conclusion of spontaneous scale symmetry breaking of scale invariant theory of Maxwell with $f$. Here $f$ is an integration constant and stands for the spontaneous breakdown. It is seen from Eq. \eqref{ac} that a confining effective potential which is Coulomb plus linear one, in the form of well-known “Cornell” potential in quantum chromodynamics (QCD) is arisen as $V(r)=-\frac{\alpha}{r}+\beta r$. Adding both the usual Maxwell term with non analytic form of $F^2$ provide the form of nonlinear electrodynamics, which is different from the $-\frac{f}{2} \sqrt{F^{2}}$ Lagranigian models using in string theory \cite{Nielsen:1973qs}. Note that $\sqrt{-F^{2}}$ is for “electrically” dominated models on the other hand  $\sqrt{F^{2}}$ is for “magnetically” dominated Nielsen-Olesen models \cite{Nielsen:1973qs}.

To solve these problems, modified gravity theories appear as alternative theories, for example, adding higher curvature corrections to the Einstein action, which is inspired by the low energy limit of string theory \cite{Nojiri:2010wj}. A vast range of modified theories and their applications to black holes and wormholes now exist in the literature \cite{Jusufi:2017mav,Jusufi:2017lsl}. Since the Lovelock gravity and its simplest case Einstein-Gauss-bonnet (EGB) gravity are the most well-known theories of modified gravity \cite{Lovelock:1972vz,Boulware:1985wk}, there were many studies on these theories for $D > 4$ black holes \cite{Wiltshire:1988uq}, on the other hand for $D=4$, EGB gravity, is topological invariant, has not any physical contribution to dynamics \cite{Simon:1990ic}. Based on non-Lagrangian approach \cite{Tomozawa:2011gp,Glavan:2019inb} by rescaling the Gauss–Bonnet dimensional coupling constant $\alpha$ as $(D-4)\alpha \to \alpha$, it was shown that it is possible to construct a new theory at limit $D \to 4$ of solutions without singularities, which is free from Ostrogradsky instability as well as bypass the Lovelock’s theorem. In addition, the GB term contributes to GR as a quantum correction. The properties of the EGB in $4D$ and its solutions have been studied in various papers \cite{Fernandes:2020rpa,Kruglov:2021stm,Kumar:2020owy,Guo:2020zmf,Konoplya:2020bxa,Donmez:2020rnf,Donmez:2021fbk,Konoplya:2020qqh,Ghosh:2020syx,Singh:2020mty}. On the other hand, EGB in 4D was debated in literature \cite{Arrechea:2020evj,Gurses:2020rxb,Gurses:2020ofy,Arrechea:2020gjw,Cao:2021nng,Aoki:2020iwm,Aoki:2020lig,Aoki:2020ila,Hennigar:2020lsl,Lu:2020iav,Fernandes:2020nbq,Fernandes:2021dsb}. G\"urses et al. claim that the EGB theory lacks of the continuity and covariant description at 4D \cite{Gurses:2020rxb}, similarly Arrechea et al. show the divergent contributions to the field equations in 4D \cite{Arrechea:2020gjw}. However, other authors claim that the problems disappear in certain circumstances and the theory is (locally) conformally flat, but the theory in 4D has preferred spacetimes, and it is not diffeomorphism invariant \cite{Cao:2021nng}.  Some authors recognize the problems but also argue that the theory can be formulated in a consistent way to cosmology and gravitational waves \cite{Aoki:2020iwm}, and show the theory either breaks the diffeomorphism invariance or has additional degrees of freedom in line with the Lovelock theorem. Moreover, the dimensional regularization of \cite{Glavan:2019inb} is made righteous in the sight of some class of spherically symmetric spacetime metric that typically present a new scalar degree of freedom \cite{Aoki:2020lig,Aoki:2020ila,Hennigar:2020lsl,Lu:2020iav}. Moreover, it is shown that the field equation for the EGB is intimately connected with generalized conformal properties of the scalar field \cite{Fernandes:2020nbq,Fernandes:2021dsb}. Aoki et al. study the gravitational waves (GW) in EGB and find a bound for the
EGB parameter as $\tilde{\alpha} \lesssim \mathcal{O}(1) \mathrm{eV}^{-2}$ \cite{Aoki:2020iwm}, on the other hand using the velocity propagation of GW, Clifton el al. show the bound at $\alpha \approx 10^{49} \mathrm{eV}^{-2}$ \cite{Clifton:2020xhc}. Hence, the spherically symmetric black hole solutions in the 4DEGBG are still valid in \cite{Glavan:2019inb} and therefore, it is worth studying more features of the black hole solutions in regularized 4DEGBG \cite{Fernandes:2020nbq,Hennigar:2020lsl}.

Main aim of the study is to find a possible new effects by coupling the confining potential generating nonlinear gauge field system in the scalar-tensor description of regularized 4DEGBG black hole. Here, we mainly attend to a regularized 4DEGB theory which is obtained in Ref. \cite{Fernandes:2020nbq,Hennigar:2020lsl}, that present a counter-term to eliminate the divergent part of the theory. In 1992 Mann also used same procedure in 2 dimensions \cite{Mann:1992ar}, which construct a well defined theory, costing only adding new scalar degree of freedom.

This paper is organized as follows. Section 2 contains the brief review of scalar-tensor description of regularized 4DEGBG, and then we solve the theory and present an exact black hole solution for the scalar-tensor description of regularized 4DEGBG with confining electric potential. In Section 3, we study some basic thermodynamics properties of the black hole, discuss their properties, and give numerical analyses. In Section 4, shadow of the black hole, and in Section 5, quasinormal modes frequencies are discussed. Finally, a conclusion is presented in Sect. 6.

\section{Black hole with Confining Electric Potential in regularized 4DEGBG}

The action of the scalar-tensor formulation of regularized 4DEGBG coupled to non-linear gauge theory containing $\sqrt{-F^{2}}$ term is written as follows \cite{Hennigar:2020lsl,Lu:2020iav,Kobayashi:2020wqy,Mann:1992ar,Fernandes:2020nbq,Fernandes:2021dsb,Fernandes:2021ysi}:
 
\begin{equation}
S=\frac{1}{16\pi} \int_{\mathcal{M}} d^{4} x \sqrt{-g}\Big[R+\alpha \big(4 G^{\mu \nu} \nabla_{\mu} \phi \nabla_{\nu} \phi-\phi \mathcal{G}+4 \square \phi(\nabla \phi)^{2}+2(\nabla \phi)^{4}\big) 
+\mathcal{L}_{NGF}\Big]   \,.  \label{action}
\end{equation}%
Note that $g$ is the determinant of the metric tensor $g_{ab}$, $\alpha$ is a dimensionless coupling constant, and $\mathcal{L}_{N G F}$ is the
Lagrangian for the nonlinear gauge field. On the other hand, $\mathcal{G}$ stands for the Gauss-Bonnet invariant and $\phi$ is a scalar field. The action in Eq. \ref{action} can be derived from the Lovelock theory $\mathcal{L} = \sqrt{-g} \left( - 2 \Lambda +R + \alpha \mathcal{G} + \dots \right) \, ,$ with $
\mathcal{G} \equiv R^2 - 4 R_{\mu \nu} R^{\mu \nu} + R_{\alpha \beta \mu \nu} R^{\alpha \beta \mu \nu} \,,$
using the by the addition of a counter-term that consists of the Gauss-Bonnet invariant of a conformally transformed geometry $\tilde{g}_{\mu \nu}=e^{2\phi} g_{\mu \nu}$. Then it gives \cite{Fernandes:2020nbq,Hennigar:2020lsl}
\begin{equation}
 \lim_{D \to 4} \frac{\int_{\mathcal{M}} d^{D} x \sqrt{-g} \mathcal{G}-\int_{\mathcal{M}} d^{D} x \sqrt{-\tilde{g}} \tilde{\mathcal{G}}}{D-4} =     \int_{\mathcal{M}} d^{4} x \sqrt{-g} \br{4 G^{\mu \nu} \nabla_{\mu} \phi \nabla_{\nu} \phi-\phi \mathcal{G}+4 \square \phi(\nabla \phi)^{2}+2(\nabla 
 \phi)^{4}} \, ,
\end{equation}

where the factor of $D-4$ is appeared due to regularization of  $\alpha\rightarrow\alpha/(D-4)$ for observational constraints $\alpha >0$ \cite{Clifton:2020xhc} which is different than the method of Glavan \& Lin. Here it is shown that the counter-term has tildes, to remove divergences in the 4 dimensional limit. Hence, there is a well-defined scalar-tensor theory of gravity in (\ref{action}).

We vary the action Eq.~(\ref{action}) with respect to metric, and obtain the following equations of motions (EOMs):
\begin{equation} \label{feqs0}
    G_{\mu \nu} + \alpha \mathcal{H}_{\mu \nu}=8\pi \, T_{\mu \nu}\, ,
\end{equation}
with the energy-momentum tensor $T_{\mu \nu}$ and
\begin{equation} \label{feqs}
\begin{aligned}
\mathcal{H}_{\mu\nu} =& 2G_{\mu \nu} \dpp+4P_{\mu \alpha \nu \beta}\left(\nabla^\beta \nabla^\alpha \phi - \nabla^\alpha \phi \nabla^\beta \phi\right) +4\left(\nabla_\a \phi \nabla_\mu \phi - \nabla_\alpha \nabla_\mu \phi\right) \left(\nabla^\a \phi \nabla_\nu \phi - \nabla^\a \nabla_\nu \phi\right)\\
&+4\left(\nabla_\mu \phi \nabla_\nu \phi - \nabla_\nu \nabla_\mu \phi\right) \dal +g_{\mu \nu} \Big(2\left(\dal\right)^2 - \left( \nabla \phi\right)^4 + 2\nabla_\b \nabla_\a\phi\left(2\nabla^\a \phi \nabla^\b \phi - \nabla^\b \nabla^\a \phi \right) \Big)\,,
\end{aligned}
\end{equation}
where the double dual of the Riemann tensor is 
\begin{equation}
P_{\alpha \beta \mu \nu} \equiv \frac{1}{4} \epsilon_{\alpha \beta \gamma \delta} R^{\rho \sigma \gamma \delta} \epsilon_{\rho \sigma \mu \nu} = 2\, g_{\alpha [\mu}G_{\nu] \beta} + 2\, g_{\beta [\nu} R_{\mu] \alpha} -R_{\alpha \beta \mu \nu}.
\end{equation}
Note the square brackets stands for anti-symmetrization. 

We vary the action Eq.~(\ref{action}) with respect to $\phi$, and obtain the following equations of motions:
\begin{equation} \label{scalareq}
\begin{aligned}
&R^{\mu \nu} \nabla_{\mu} \phi \nabla_{\nu} \phi - G^{\mu \nu}\nabla_\mu \nabla_\nu \phi - \dal \dpp +(\nabla_\mu \nabla_\nu \phi)^2 
- (\dal)^2 - 2\nabla_\mu \phi \nabla_\nu \phi \nabla^\mu \nabla^\nu \phi = \frac{1}{8}\mathcal{G} \, .
\end{aligned}
\end{equation}
This equation is conformal invariance under the transformation $g_{\mu \nu} \to g_{\mu \nu} e^{2\sigma}$ and $\phi \to \phi -\sigma$ \cite{Fernandes:2021dsb}. Moreover,  using Eq. (\ref{scalareq}), with \eqref{feqs0}, it becomes \cite{Fernandes:2020nbq}
\begin{equation}
    R+\frac{\alpha}{2}\mathcal{G} = -8\pi \,T.
    \label{Eq:trace}
\end{equation}
Hence Fernandes shows that this purely geometric relation is a direct consequence of the conformal invariance of the scalar field equation \cite{Fernandes:2021dsb}. Then one can calculate a Noether current with vanishing divergence \cite{Saravani:2019xwx}:
\begin{equation}
    j^{\mu} = \frac{1}{\sqrt{-g}}\frac{\delta S}{\delta (\partial_\mu \phi)}, \qquad {\rm such\,\, that} \qquad \nabla_\mu j^\mu = 0 \,. \label{eq:current}
\end{equation}
Note that one can vanish divergence $\nabla_\mu j^\mu = 0$ to show that $\partial_\mu \left(\sqrt{-g} j^\mu\right) = 0$, and then find the equation of motion (\ref{scalareq}). 

Using the field equations (\ref{feqs0}) and (\ref{scalareq}), we will construct the black hole solution with confining electric potential in the regularized 4DEGBG gravity.  To construct the black hole, we consider the $4$ dimensional
static and spherically symmetric metric: 
\begin{equation}
ds^{2}=-A(r)dt^{2}+\frac{1}{A(r)}dr^{2}+r^2\left(d\theta^2 + \sin^2 \theta d\varphi^2\right),
\end{equation}

On the other hand, Lagrangian for the nonlinear gauge field is given by \cite{Vasihoun:2014pha}:

\begin{equation}
L_{NGF}=-\frac{1}{4}F^{2}-\frac{f}{2}\sqrt{-F^{2}},
\end{equation}

where $F^{2}\equiv F_{\kappa \lambda }F_{\mu \nu }g^{\kappa \mu }g^{\lambda
\nu }.$ This model  produces a confining effective potential which is $V=-%
\frac{a}{r}+\beta r$ of the form of the well-known Cornell potential in quantum chromodynamics (QCD) \cite{Korover:2009zz,Gaete:2006xd}. It is crucial to stress that
the Lagrangian contains both the usual Maxwell term as well as a non-analytic function of $F^{2}$ and thus it is a non-standard form of nonlinear electrodynamics. 
The energy-momentum tensor of the nonlinear gauge field is \cite{Vasihoun:2014pha,Guendelman:2003ib,Gaete:2006xd,Korover:2009zz,Guendelman:2011sm,Guendelman:2012ve,Guendelman:2018tzi}:

\begin{equation}
T_{\mu \nu }^{(NGF)}=\left( 1-\frac{f}{\sqrt{-F^{2}}}\right) F_{\mu \kappa
}F_{\nu \lambda }g^{\kappa \lambda }-\frac{1}{4}\left( F^{2}+2f\sqrt{-F^{2}}%
\right) g_{\mu \nu },
\end{equation}

and

\begin{equation} \label{eom1}
\partial_{\nu}\left(\sqrt{-g}\left(1-\frac{f}{\sqrt{-F^{2}}}\right)
F_{\kappa \lambda} g^{\mu \kappa} g^{\nu \lambda}\right)=0.
\end{equation}

In this case the gauge field equations of motion \eqref{eom1} become: 
\begin{equation}
\partial _{r}\left( r^{2}\left( F_{0r}-\frac{\varepsilon _{F}f}{\sqrt{2}}%
\right) \right) =0\quad ,\quad \varepsilon _{F}\equiv sign\left(
F_{0r}\right) 
\end{equation}

in which 
\begin{equation}
F_{\mu \nu }=0\text{ for }(\mu ,\nu )\neq (0,r)\quad ,\quad F_{0r}=F_{0r}(r)
\end{equation}%
and its solution reads: 
\begin{equation}
F_{0r}=\frac{\varepsilon _{F}f}{\sqrt{2}}+\frac{Q}{\sqrt{4\pi }r^{2}}\quad
,\quad \varepsilon _{F}=sign(Q).
\end{equation}

Again, as in the flat space-time case \eqref{ac}, 
the electric field contain a
radial constant piece ${\varepsilon _{F}f}\sqrt{2}$ alongside with the Coulomb term. \ The form of a QCD-like (Cornell -type) confining-type potential (provided $\varepsilon _{F}f<0$ )with $\varepsilon _{F}$ . Then the energy momentum tensor for the nonlinear gauge field is

\begin{equation} \label{conf}
T_{0}^{0}=T_{\eta}^{(F)}{\eta}=-\frac{1}{2} F_{0 \eta}^{2} \quad, \quad T_{i j}=g_{i j}\left(\frac{1}{2} F_{0 \eta}^{2}-\frac{f}{\sqrt{2}}\left|F_{0 \eta}\right|\right).
\end{equation}

After we solve the field equations Eq. \ref{conf} within  Eq. \ref{feqs0}:
\begin{eqnarray}
 &\frac{\left[2 \alpha  (1-A(r))+r^2\right] A'(r)}{r^3}-\frac{[1-A(r)] \left(r^2-\alpha  (1-A(r))\right)}{r^4}  +4\pi\left(\frac{f}{\sqrt{2}}-\frac{Q}{2 \sqrt{\pi } r^2}\right)^2
 =0.
\end{eqnarray}


with $M$ the ADM mass, we get
\begin{widetext}
\begin{equation}
A(r)=1+\frac{r^{2}}{2\alpha }\left( 1\pm \sqrt{1+4\alpha \left(\frac{2 \pi  f^2}{3}-\frac{2 \sqrt{2 \pi } f Q}{r^2}+\frac{2 M}{r^3}-\frac{Q^2}{r^4}\right) }\right),  \label{fr}
\end{equation}
\end{widetext}
with the scalar field profile for this solution is 
\begin{equation}
\phi'(r)=\frac{1-\sqrt{A(r)}}{r\sqrt{A(r)}}.
\end{equation}
Note that the prime stands for a derivative with respect to $r$. The $\pm$ sign in front of the square root term in Eq. \eqref{fr}, corresponds to two different branches of solutions. It is shown that the negative branch of solution is for a charged 4D black hole, on the other hand, the positive branch gives instabilities of the graviton, because of the positive sign in the mass \cite{Singh:2020mty}. Thus, since only the negative branch leads to a
physically meaningful solution, we will limit our analysis
to this branch of the solution. Note that the metric function reduces to the scalar-tensor description of regularized 4DEGBG solution at the limit
of $f=Q=0$ founded by Fernandes et al. \cite{Fernandes:2020nbq}. Moreover, when a
confining-type charge term equals to zero $f=0,$ it reduces
to the charged EGBG\ black hole solution founded by Fernandes \cite{Fernandes:2020rpa}.

The asymptotic of the
metric function $A(r)$ when $r \to \infty$ is

\begin{equation}
A(r)=1-\,{\frac {2M}{r}}+{\frac {{Q}^{2}}{{r}^{2}}}+\frac{2}{3}{r}^{2}\pi\,{f}^{2
}+2\sqrt{2 \pi}f Q+\mathcal{O}\left(M^{2},Q^{3},f^{3}\right).
\end{equation}

\begin{figure}[ht!]
   \centering
   \includegraphics[scale=0.6]{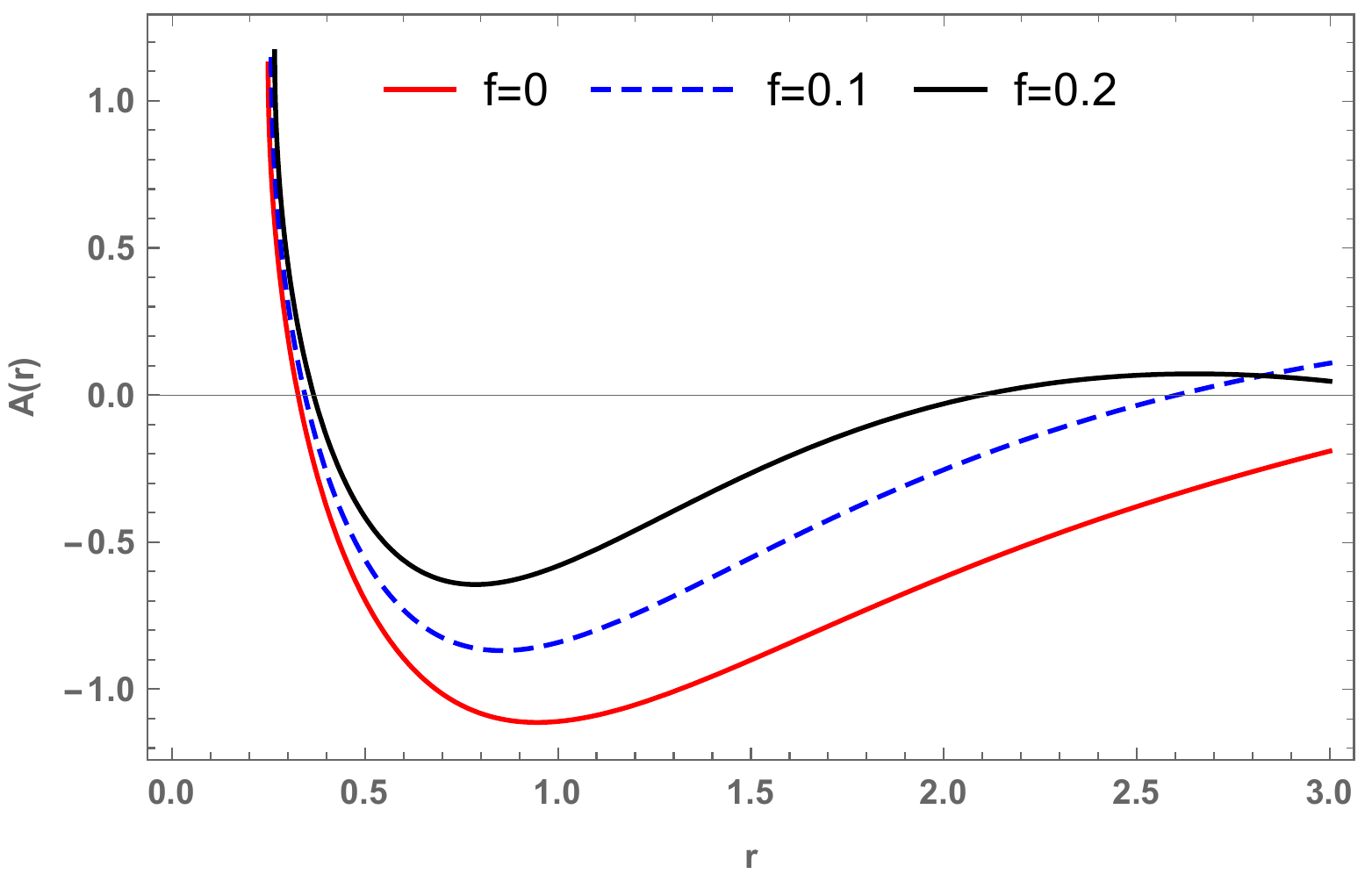}
    \caption{The lapse function $A(r)$ as a function of $r$ for $M=2$, $Q=1$, $\alpha=+0.2$ and for the different values of $f$.}
    \label{fig:hr1}
\end{figure}

 The event horizon of the black holes $r_{+}$ is the larger root of the  equation Eq. \eqref{fr}, $A(r)=0$. As shown in Fig. \eqref{fig:hr1}, the number of horizons depends on the parameters $f$.

\section{Thermodynamics and physical properties}

\begin{figure}[ht!]
   \centering
    \includegraphics[scale=0.6]{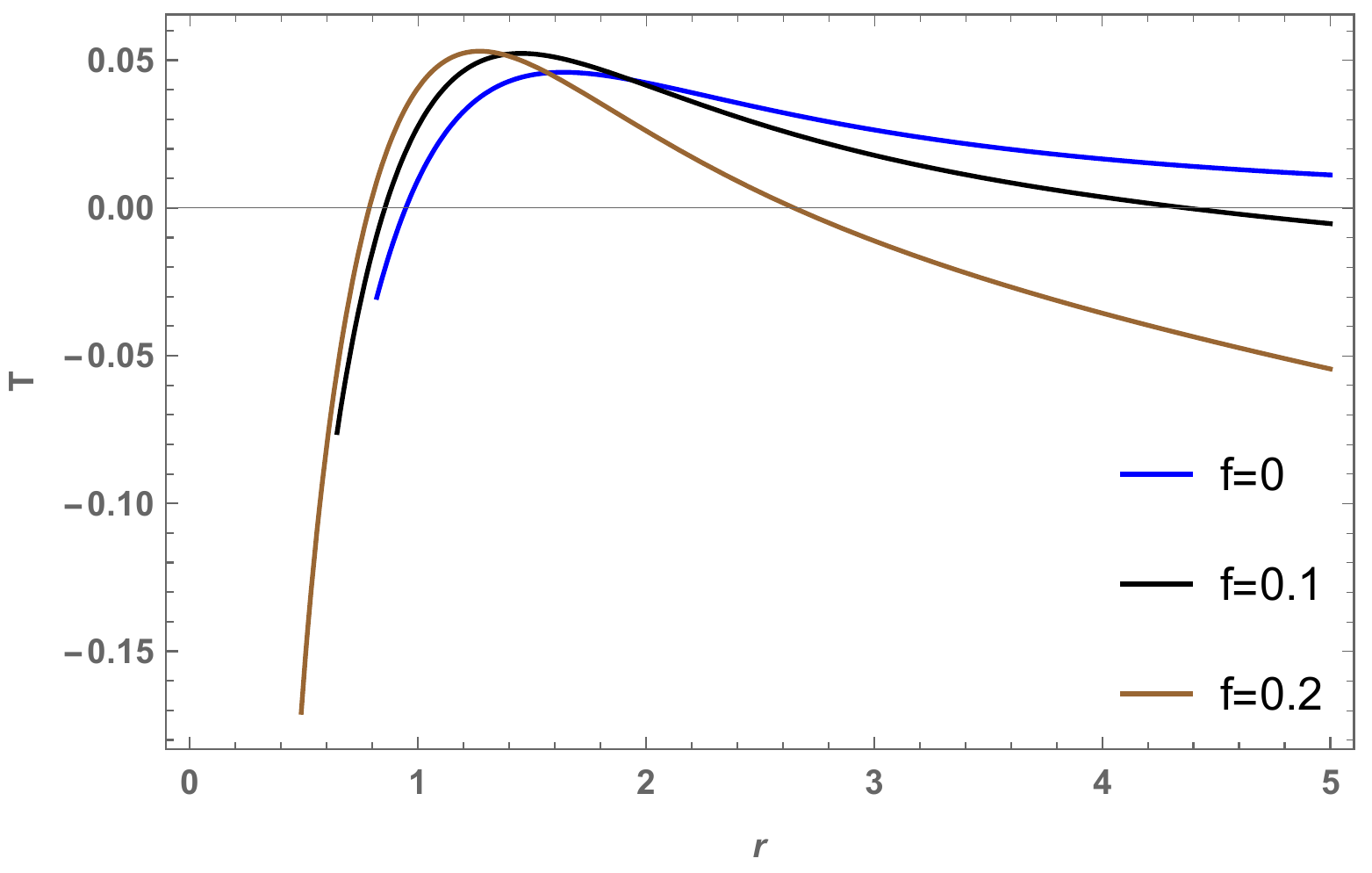}
    \caption{Hawking temperature $T$ versus $r$ for $M=2$, $Q=1$, $\alpha=+0.2$ and for the different values of $f$.}
    \label{fig:temp}
\end{figure}

 In this section, we study some thermodynamic properties of the black hole. 
First, the black hole mass can be calculated by $\left.g_{00}\right|_{r=r_{+}}=0$
\begin{equation}
M(r_{+})= \frac{3 \alpha -2 \pi  f^2 r_{+}^4+6 \sqrt{2 \pi } f Q r_{+}^2+3 Q^2+3 r_{+}^2}{6 r_{+}}.
\end{equation}
Note that one can take a limit of vanishing Gauss-bonnet coupling parameter, and charges $\alpha=Q=f=0$, then the mass of the black hole reduces to mass of Schwarzschild
black holes: $M_{+}=\frac{r_{+}}{2}$. Second, the Hawking temperature associated with horizon radius $r_{+}$, can be obtained through:
\begin{widetext}

\begin{equation}T_{+}=\frac{\kappa}{2 \pi}=\frac{\frac{r_{+} \left(1-\sqrt{\frac{4 \alpha  \left(2 \pi  f^2 r_{+}^4-6 \sqrt{2 \pi } f Q r_{+}^2+6 M r_{+}-3 Q^2\right)}{3 r_{+}^4}+1}\right)}{\alpha }-\frac{2 \left(2 \sqrt{2 \pi } f Q r_{+}^2-3 M r_{+}+2 Q^2\right)}{r_{+}^3 \sqrt{\frac{8}{3} \pi  \alpha  f^2-\frac{8 \sqrt{2 \pi } \alpha  f Q}{r_{+}^2}+\frac{8 \alpha  M}{r_{+}^3}-\frac{4 \alpha  Q^2}{r_{+}^4}+1}}}{4 \pi },\end{equation}
\end{widetext}
where $\kappa$ is the surface gravity given by

\begin{equation} \kappa=\left.\sqrt{-\frac{1}{2} \nabla_{\mu} \chi_{\nu} \nabla^{\mu} \chi^{\nu}} \equiv \frac{1}{2} \frac{\partial A(r)}{\partial r}\right|_{r=r_{+}},\end{equation}
and the Hawking temperature is plotted in Fig. \eqref{fig:temp}. Furthermore, we can take a limit of vanishing Gauss-bonnet coupling parameter, and charges $\alpha=Q=f=0$, then the Hawking temperature of the black hole reduces to temperature of Schwarzschild
black holes: $T_{r_{+}}=\frac{1}{8M\pi}$.

\begin{figure}[ht!]
   \centering
    \includegraphics[scale=0.6]{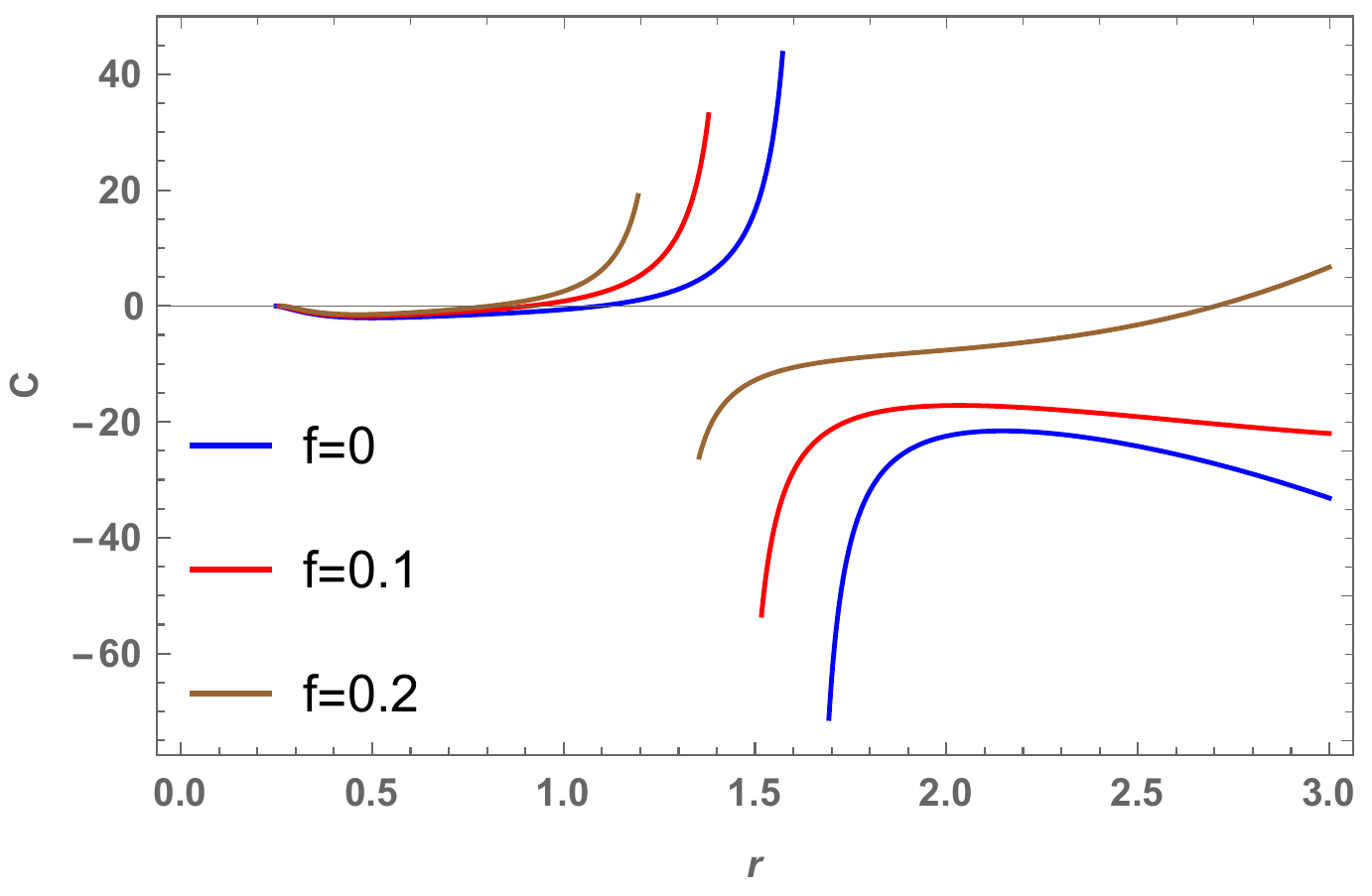}
    \caption{Specific heat $C_{+}$ versus horizon $r_{+}$ for $M=2$, $Q=1$, $\alpha=+0.2$ and for the different values of $f$.}
    \label{fig:spec}
\end{figure}

Moreover, to check the thermodynamics stability of black holes, the
behaviour of specific heat $(C_+)$ of the black holes is calculated. The positive (negative) specific heat
signifies the local thermodynamics stability (instability) of the black holes. By using the
relation $C_{+}=\frac{\partial M_{+}}{\partial r_{+}} / \frac{\partial T_{+}}{\partial r_{+}}$, the specific heat is plotted in Fig. \eqref{fig:spec}. It is shown that the regularized 4DEGBG black hole with confining electric potential is thermodynamically stable, where the Hawking temperature in Fig. \eqref{fig:temp} and the specific heat in Fig. \eqref{fig:spec} are positive, in some range of confining charge parameters and event horizon radii.

\section{Shadow of the black hole with confining electric potential in regularized 4DEGBG}
Studying the shadow of the black hole is recently gained more attention, (see, for example, \cite{Falcke:1999pj,Cunha:2018acu,Atamurotov:2013sca,Bisnovatyi-Kogan:2018vxl,Cunha:2016wzk,Vagnozzi:2019apd,Shaikh:2018lcc,Cunha:2017eoe,Tsukamoto:2014tja,Ovgun:2020gjz,Ovgun:2018tua} and references therein). Therefore, now we analyse the shadow of the black hole and study the effect of the confining charge on the shadow cast. The Hamilton-Jacobi approach  for a photon in the equatorial plane $\theta=\frac{\pi}{2}$ can be written in the form of

\begin{equation}
H=\frac{1}{2} g^{\mu v} p_{\mu} p_{v}=\frac{1}{2}\left(\frac{L^{2}}{r^{2}}-\frac{E^{2}}{A(r)}+\frac{\dot{r}^{2}}{A(r)}\right)=0. \label{hj}
\end{equation}
Note that $p_{\mu}$is the photon momenta, $\dot{r}=\partial H / \partial p_{r}$ , $E = -p_t$ is the energy and $L = p_\phi$ is the angular momentum. Using the \eqref{hj}, a complete description of the dynamics with effective potential $V$ is given by
\begin{equation}
V+\dot{r}^{2}=0, \quad V=A(r)\left(\frac{L^{2}}{r^{2}}-\frac{E^{2}}{A(r)}\right) .
\end{equation}
The stability condition of the circular null geodesics provides that $V(r)=V'(r)=0$ and $V''(r)>0$ \cite{Ovgun:2020gjz}. For the circular photon orbits, unstability associated to the maximum value of the effective potential so that 

\begin{equation}
\left.V(r)\right|_{r=r_{p}}=0,\left.\quad V'(r)\right|_{r=r_{p}}=0,
\end{equation}
where the impact parameter $b \equiv \frac{L}{E}=\frac{r_{p}}{\sqrt{A\left(r_{p}\right)}}$ and $r_p$ is the radius of the photon sphere, which can be calculated by finding the largest root of the this relation:
\begin{equation} \label{phsp}
\frac{A^{\prime}(r_p)}{A(r_p)}=\frac{2}{r_p},
\end{equation}

The above equation \ref{phsp} is complicated to solve analytically, just so, the numerical methods are used to obtain the photon sphere radius $r_p$ which are presented
in Table I. It is shown that increasing values of the confining charge parameter tend to increase the photon sphere.

The radius of the black hole shadow as observed
by a static observer at the position $r_0$ is
\begin{equation}
R_{s}=r_{p} \sqrt{\frac{A\left(r_{0}\right)}{A\left(r_{p}\right)}},
\end{equation}
and for large distant observer ($A(r_{0})=1$) and 

\begin{equation} \label{shadow}
R_{s}^2=\frac{ r_p^{2}}{A(r_p)}.
\end{equation}

	\begin{table}[ht!]
    \centering
    \begin{tabular}{ |p{1cm}||p{2cm}|p{2cm}| }
    \hline
        $f$ &  $r_{p}$ &  $R_{s}$ \\ [0.5ex] 
        \hline
        0.20 & 2.52725& 9.65169  \\
        0.30 & 1.89395& 8.88559 \\
         0.40 &1.4573& 4.68954 \\
         0.50 & 1.1166 & 2.62366 \\
        \hline
    \end{tabular}
    \caption{Effects of the confining charge on the BH shadow for fixed $M=2$, $Q=1$ and $\alpha=0.2$.}
    \label{table1}
\end{table}

\begin{figure}[ht!]
   \centering
    \includegraphics[scale=0.5]{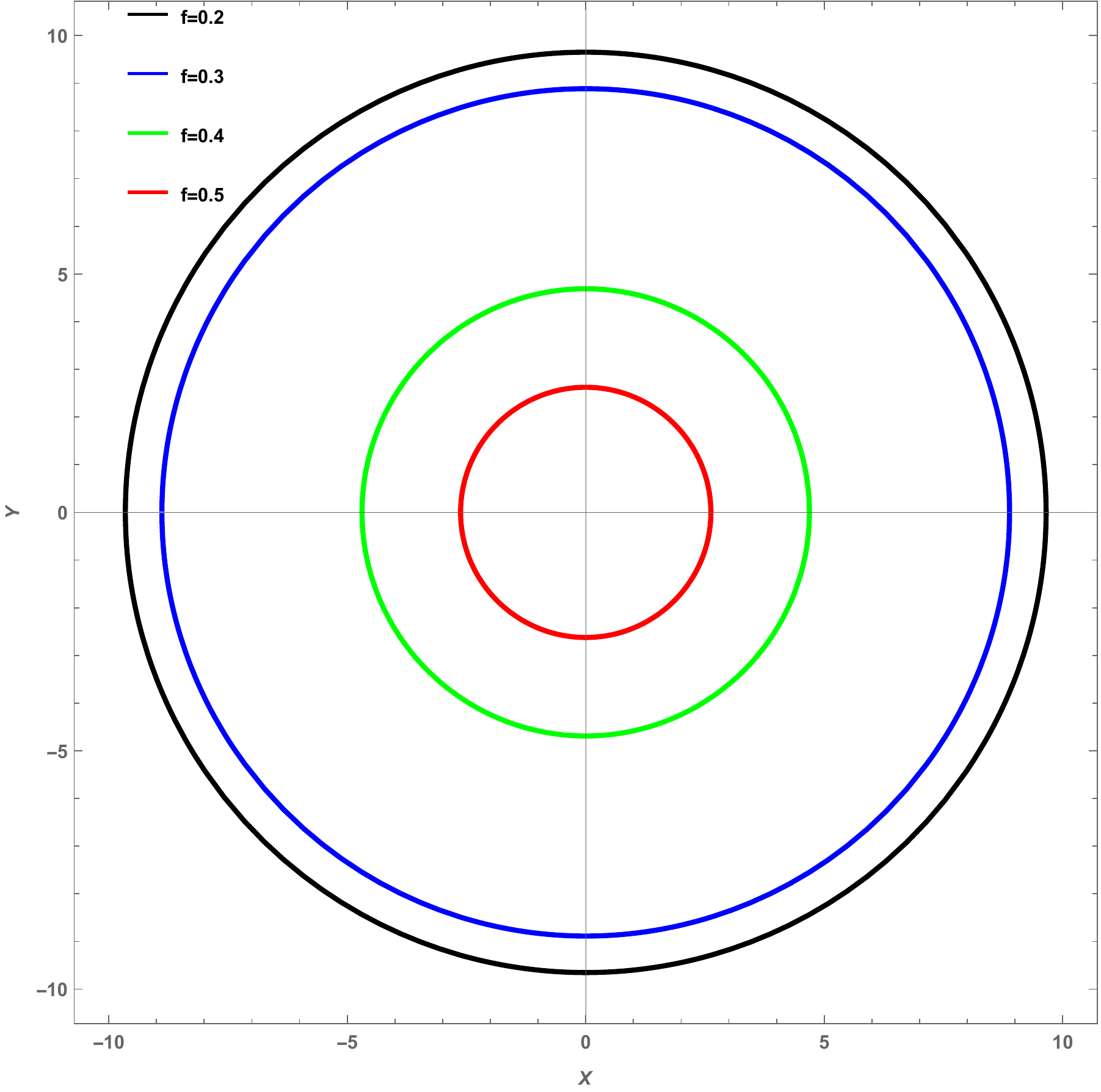}
    \caption{Shadow of the black hole for different values of $f$ for $M=2$, $Q=1$, $\alpha=+0.2$.}
    \label{fig:shadow}
\end{figure}

The apparent shape of a shadow is plotted by a stereographic projection in terms of the celestial coordinates $X$ and $Y$ in Fig. \eqref{fig:shadow}, that the radius of the shadow increases with the decreasing value of $f$. Fig. \eqref{fig:shadow} shows the evidence that confining charge $f$
has a stronger effect on the shadow size of the 4D EGBG black hole.

\begin{figure}[ht!]
   \centering
    \includegraphics[scale=0.4]{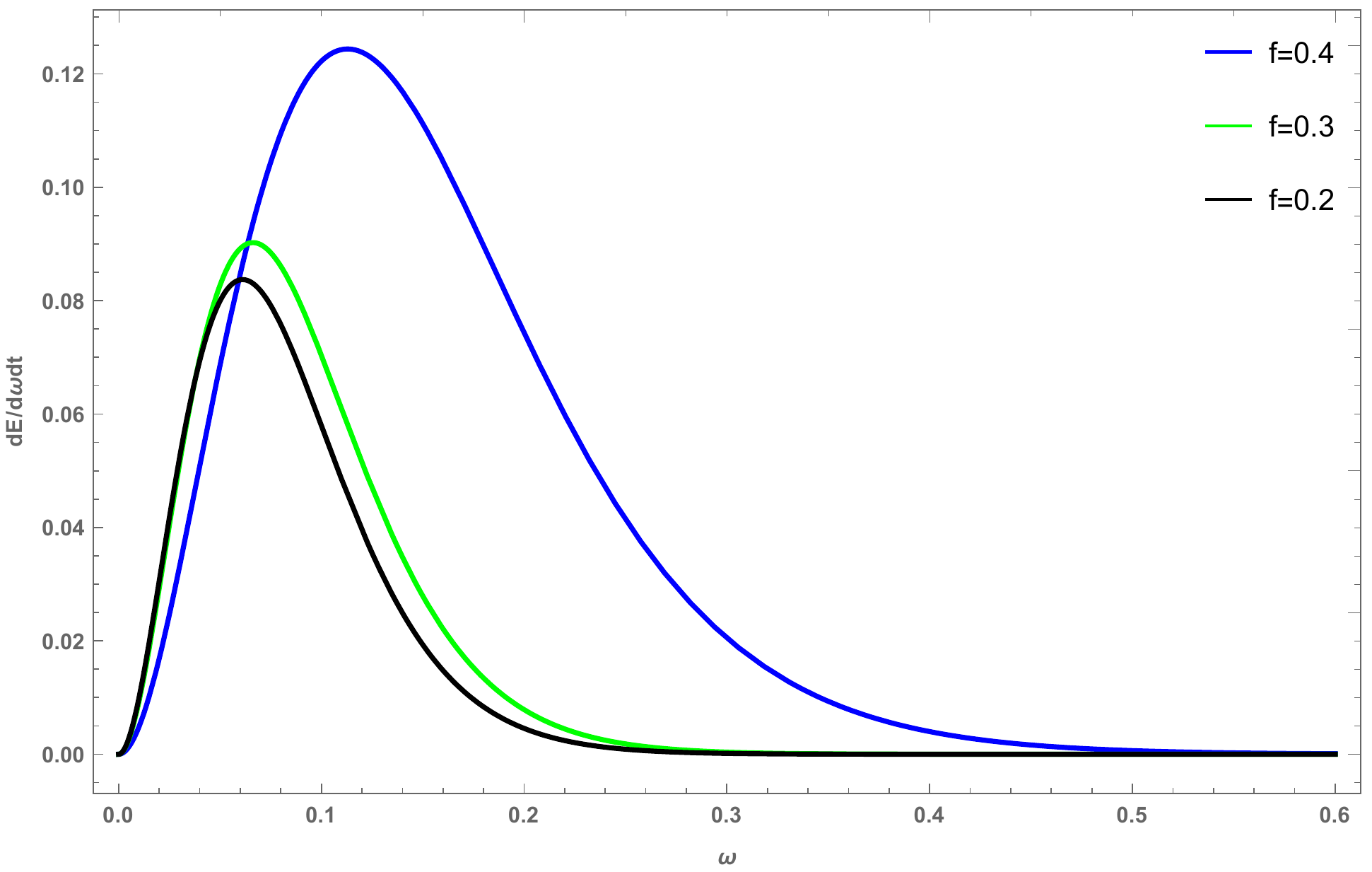}
    \caption{Energy emission rate of the black hole for different values of $f$ for $M=2$, $Q=1$, $\alpha=+0.2$.}
    \label{fig:em}
\end{figure}

The relation between the high energy absorption cross section and the shadow for the observer located at infinity is 
\begin{equation}
\frac{d^{2} E(\omega)}{d t d \omega}=\frac{2 \pi^{3} \omega^{3} R_{s}^{2}}{\exp \left(\omega / T_{+}\right)-1},
\end{equation} where $\omega$ stands for the emission frequency. In Fig. \eqref{fig:em}, indicates that the there is a peak of the energy emission rate and increasing the value of $f$, the peak of the energy emission rate increases.

\section{Quasinormal modes}
\begin{figure}[ht!]
   \centering
    \includegraphics[scale=0.6]{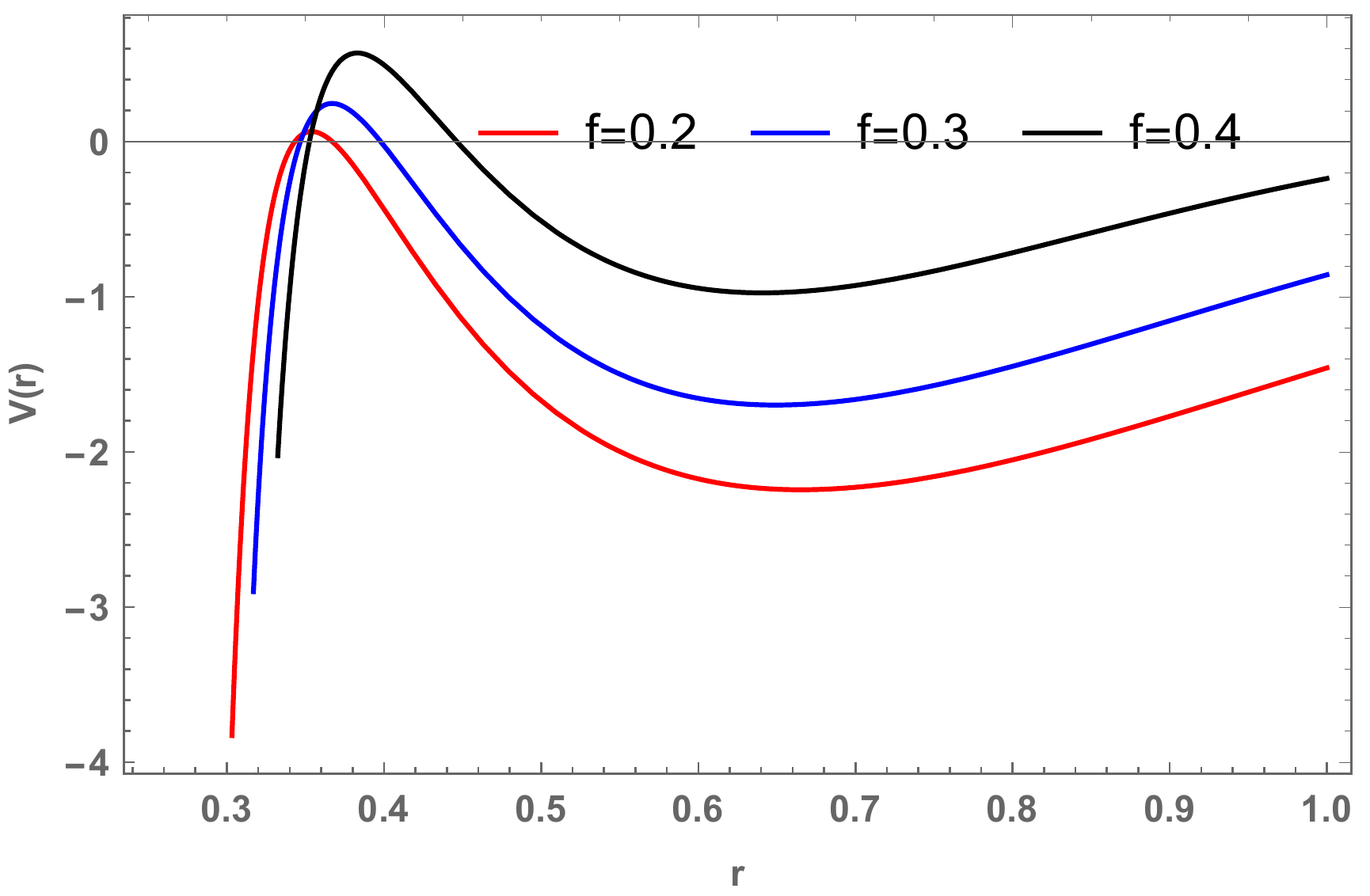}
    \caption{Effective potential for different values of $f$ for $M=2$, $Q=1$, $l=1$, and $\alpha=+0.2$.}
    \label{fig:Veff}
\end{figure}

Since the discovery of the LIGO experiment, gravitational waves which is emitted by perturbed black holes become popular, that are dominated by `quasinormal ringing'. Quasinormal modes are damped oscillations at single frequencies which are characteristic of the underlying system and can be calculated using the scalar perturbation of massless field around black hole: 

\begin{equation} \label{scalar}
\frac{1}{\sqrt{-g}} \partial_{\mu}\left(\sqrt{-g} g^{\mu v} \partial_{\nu} \Phi\right)=0,
\end{equation}
and separate the above equation using the \begin{equation}
\Phi=e^{-i \omega t} \frac{R(r)}{r} e^{i m \phi} Y_{l m}(\vartheta),
\end{equation}
where separation of variables for radial function $R(r)$ and spherical harmonics $Y_{l m}$. Using the tortoise coordinates $
r_{*}=\int d r / f(r)$,  the Eq. \ref{scalar} for the radial field becomes: 
\begin{equation}
\frac{d^{2} R\left(r_{*}\right)}{d r_{*}^{2}}+\left(\omega^{2}-V\left(r_{*}\right)\right) R\left(r_{*}\right)=0,
\end{equation}
with the effective potential ($l=1,2,3$ is multipole number, and $\omega=\omega_{R}-i \omega_{I}$ is a complex quasinormal mode frequency)
\begin{equation}
V(r)=A(r)\left(\frac{A^{\prime}(r)}{r}+\frac{l(l+1)}{r^{2}}\right).
\end{equation}
which is plotted in Fig. \eqref{fig:Veff}. 

Here we use the WKB approximation method to find quasinormal mode frequency and the unstable circular null geodesic method \cite{Cardoso:2008bp,Konoplya:2017wot}.
The imaginary part of the quasinormal mode frequency (Im $\omega$=-$\omega_I$) which is responsible for the temporal, exponential decay can be calculated , in the
large-$l$ limit as follows:

\begin{equation}
\omega_{l \gg 1}=l \Omega_{p}-i\left(n+\frac{1}{2}\right)\left|\lambda_{\mathrm{L}}\right|, \end{equation}
with the angular velocity $\Omega_{p}$:
\begin{equation}
\Omega_{p}=\frac{\sqrt{A\left(r_{p}\right)}}{r_{p}}, \end{equation} 
and Lyapunov exponent $\lambda_{\mathrm{L}}$:
\begin{equation}
\lambda_{\mathrm{L}}=\sqrt{\frac{A\left(r_{p}\right)\left[2 A\left(r_{p}\right)-r_{p}^{2} A^{\prime \prime}\left(r_{p}\right)\right]}{2 r_{p}^{2}}},
\end{equation}
where $n$ is the overtone number and take values $n= 0, 1, 2,...$. Table \ref{tab:table2}  shows the real part as well as imaginary part of the quasinormal modes increasing with the increasing $f$. Hence, the modes are stable because in the Table \ref{tab:table2} the the imaginary
parts of the quasinormal modes frequencies (Im $\omega$=-$\omega_I$) are negative, furthermore, the real part stands for the frequency of oscillations. Explained differently, increasing
the confining charge parameter $f$, the scalar perturbations oscillate with greater frequency $\omega$ which means that oscillates decay faster.

	\begin{table}[ht!]
    \centering
    \begin{tabular}{ |p{1cm}||p{2cm}|p{2cm}||p{2cm}| }
    \hline
        $f$ &  $\omega_{R}$ & $\omega_{I}$ \\ [0.5ex] 
        \hline
        0.20 & 0.103609 & 0.0641446 \\
        0.30 & 0.112542 & 0.0723644 \\
        0.40 & 0.213241 & 0.134316  \\
         0.50 & 0.381146 & 0.209336  \\
        \hline
    \end{tabular}
    \caption{Effects of the confining charge on the quasinormal modes frequencies for fixed $M=2$, $Q=1$, $n=0$, $l=1$ and $\alpha=0.2$.}
    \label{tab:table2}
\end{table}

\section{Conclusion}

In this paper, we obtain the exact BH solution in the scalar-tensor description of regularized 4DEGBG including confining electric charge. EGB gravity theory is known as nontrivial for in $D>4$,  notwithstanding, till one shows that dimensional reduction by re-scaling of the coupling constant $\alpha$ as $(D-4)\alpha \to \alpha$, open on to the novel regularization method to build the effective solutions in 4D \cite{Glavan:2019inb}. On the other hand, we show new effects of regularized 4DEGBG by coupling the confining potential which generates nonlinear gauge field system. This is also clear way to generate dynamically the cosmological constant through the non-Maxwell gauge fields \cite{Vasihoun:2014pha,Guendelman:2003ib,Gaete:2006xd,Korover:2009zz,Guendelman:2011sm,Guendelman:2012ve}. The metric function for the derived black hole solution in Eq. \eqref{fr}, reduces to the regularized 4DEGBG solution at the limit of $f=Q=0$ founded by Fernandes et al. \cite{Fernandes:2020nbq} as well as it reduces to the charged 4DEGBG black hole solution founded by Fernandes \cite{Fernandes:2020rpa} when a
confining-type charge term equals to zero $f=0$. In addition, at the limit of $\alpha \to 0$ and $Q = 0$,  the solution reduces to Schwarzschild-de-Sitter black hole with effective cosmological constant: $\Lambda_{eff}=2\pi f^2$.

Next, we study thermodynamics of the regularized 4DEGBG black hole with confining electric potential, and analyze its thermodynamics stability using the black hole mass, Hawking temperature, and specific heat of the black hole. We show that the regularized 4DEGBG with confining electric potential is thermodynamically stable, where the Hawking temperature in Fig. \eqref{fig:temp} and the specific heat in Fig. \eqref{fig:spec} are positive, in some range of confining charge parameters and event horizon radii. Moreover, the shadow size, the energy emission rate, and at last, quasinormal modes of the black hole are investigated to see the effect of confining charge on the regularized 4DEGBG black hole. The possible range of the confining charge parameter is constrained and the results indicate that increasing the parameter of confining charge shrinks the shadow's radius in Fig. \eqref{fig:shadow} and Table \ref{table1}. We show that there is a peak of the energy emission rate and increasing the value of $f$, the peak of the energy emission rate increases in Fig. \eqref{fig:em}. Furthermore, using the correspondence between null geodesics and quasinormal modes in the eikonal regime for test fields, we also observe that the increasing the parameter of confining charge, increase the real part and also imaginary part of the quasinormal modes frequencies as well as the modes are stable because in the Table \ref{tab:table2} the the imaginary
parts of the quasinormal modes frequencies (Im $\omega$=-$\omega_I$) are negative. It is worth to mention that increasing the confining charge parameter $f$, the scalar perturbations oscillate with greater frequency $\omega$ which means that oscillates decay faster. Thus, we conclude that these results showing the contribution of the NGF, mainly with confining charge, on the black holes in regularized 4DEGBG tells us clearly that, this is an important analytical solution to the regularized 4DEGBG. In future, 
we plan to extend this model to a case without spherical symmetry.

\begin{acknowledgments}
The author is grateful for helpful discussions with Eduardo I. Guendelman.
\end{acknowledgments}

\end{document}